\documentclass{aa}  

\usepackage{graphicx, color}
\usepackage{txfonts}
\usepackage{comment}
\usepackage{orcidlink}
\begin{document}

   \title{The space weather around the exoplanet GJ 436b}
   \subtitle{I. The large-scale stellar magnetic field}
   \titlerunning{The space weather around GJ 436b. I. The large-scale stellar magnetic field}

   \author{S. Bellotti \inst{1,2,3}\orcidlink{0000-0002-2558-6920}
          \and
          R. Fares \inst{4}\orcidlink{0000-0002-3301-341X}
          \and
          A. A. Vidotto \inst{3}\orcidlink{0000-0001-5371-2675}
          \and
          J. Morin \inst{5}\orcidlink{0000-0002-4996-6901}
          \and
          P. Petit \inst{1}\orcidlink{0000-0001-7624-9222}
          \and
          G. A. J. Hussain \inst{2}\orcidlink{0000-0003-3547-3783}
          \and 
          V. Bourrier \inst{6}\orcidlink{0000-0002-9148-034X}
          \and
          J.~F. Donati \inst{1}\orcidlink{0000-0001-5541-2887}
          \and 
          C. Moutou\inst{1}\orcidlink{0000-0002-2842-3924}
          \and
          \'E.~M. H\'ebrard \inst{1}
          }
   \authorrunning{Bellotti et al.}
    
   \institute{
            Institut de Recherche en Astrophysique et Plan\'etologie,
            Universit\'e de Toulouse, CNRS, IRAP/UMR 5277,
            14 avenue Edouard Belin, F-31400, Toulouse, France \\
            \email{stefano.bellotti@irap.omp.eu}
        \and
             Science Division, Directorate of Science, 
             European Space Research and Technology Centre (ESA/ESTEC),
             Keplerlaan 1, 2201 AZ, Noordwijk, The Netherlands
        \and
             Leiden Observatory, Leiden University,
             PO Box 9513, 2300 RA Leiden, The Netherlands
        \and
             Department of Physics, College of Science, 
             United Arab Emirates University, 
             P.O. Box No. 15551, Al Ain, UAE\\
             \email{rim.fares@uaeu.ac.ae}
        \and
             Laboratoire Univers et Particules de Montpellier,
             Universit\'e de Montpellier, CNRS,
             F-34095, Montpellier, France
        \and
            Observatoire Astronomique de l’Universit\'e de Gen\'eve, Chemin Pegasi 51b, 1290 Versoix, Switzerland
             }
   \date{Received ; accepted }

 
  \abstract
   {The space environment in which planets are embedded depends mainly on the host star and impacts the evolution of the planetary atmosphere. The quiet M dwarf GJ\,436 hosts a close-in hot Neptune which is known to feature a comet-like tail of hydrogen atoms escaped from its atmosphere due to energetic stellar irradiation. Understanding such star-planet interactions is essential to shed more light on planet formation and evolution theories, in particular the scarcity of Neptune-size planets below 3\,d orbital period, also known as ``Neptune desert''.}   
   {We aimed at characterising the stellar environment around GJ\,436, which requires an accurate knowledge of the stellar magnetic field. The latter is studied efficiently with spectropolarimetry, since it is possible to recover the geometry of the large-scale magnetic field by applying tomographic inversion on time series of circularly polarised spectra.}
   {We used spectropolarimetric data collected in the optical domain with Narval in 2016 to compute the longitudinal magnetic field, examine its periodic content via Lomb-Scargle periodogram and Gaussian Process Regression analysis, and finally reconstruct the large-scale field configuration by means of Zeeman-Doppler Imaging.}
   {We found an average longitudinal field of $-$12\,G and a stellar rotation period of 46.6\,d using a Gaussian Process model and 40.1\,d using Zeeman-Doppler Imaging, both consistent with the literature. The Lomb-Scargle analysis did not reveal any significant periodicity. The reconstructed large-scale magnetic field is predominantly poloidal, dipolar and axisymmetric, with a mean strength of 16\,G. This is in agreement with magnetic topologies seen for other stars of similar spectral type and rotation rate.}
   {} 


   \maketitle

%

\section{Introduction}\label{sec:introduction}

The stellar environment in which exoplanets are immersed has a significant impact on their atmospheres. Energetic phenomena associated with intense magnetic activity such as frequent flares can alter the chemical properties of the planetary atmosphere \citep{Segura2010,Guenther2020,Konings2022,Louca2023}, with hazardous consequences for habitability \citep[e.g.,][]{Tilley2019}. In particular for planets orbiting closer ($<0.1$\,au) to the host star, energetic stellar irradiation (X-rays and extreme ultra-violet) heats and expands the upper regions of the atmosphere, resulting in hydrodynamic escape \citep[e.g.,][]{Lammer2003,VidalMadjar2003,Owen2012}. In addition to the radiation from the host star, stellar particles from magnetised wind and coronal mass ejections also impact planetary atmospheres, e.g., by confining and stripping them away \citep{Carolan2021,Hazra2022}. Evaporation of planetary atmospheres is accentuated in the early stages of a planetary system \citep{Ribas2005,Allan2019,Ketzer2023}, and is one of the mechanisms proposed to explain the ``Neptune desert'', i.e. the paucity of planets with masses between 0.01 and 1\,M$_\mathrm{Jup}$ in short-distance orbits \citep[e.g.,][]{Lecavelier2007,Penz2008,Davis2009,Ehrenreich2011,Beauge2013,Lundkvist2016,Mazeh2016}. 
Likewise, photoevaporation can make a mini-Neptune lose a significant amount of hydrogen and helium, morphing it into a potentially habitable super-Earth \citep{Luger2015}.

 

\begin{table*}[!ht]
\caption{List of GJ\,436 observations collected in 2016 with Narval.}
\label{tab:log}
\centering
\begin{tabular}{lccccccr} 
\hline
Date & UT & HJD & $n_\mathrm{cyc}$ & $t_{exp}$ & S/N & $\sigma_\mathrm{LSD}$ & B$_l$\\
 & [hh:mm:ss] & [-2457464.4967] & & [s] & & [$10^{-4}I_c$] & [G]\\
\hline
Mar 16 & 23:49:19 & 0.00  & 0.00 & 4x700 & 201 & 5.7 & $-$12.3$\pm$7.0 \\
Mar 18 & 00:36:58 & 1.03  & 0.03 & 4x700 & 202 & 5.6 & $-$8.0$\pm$6.8 \\
Mar 20 & 23:25:14 & 3.98  & 0.10 & 4x700 & 271 & 4.0 & $-$4.0$\pm$5.0 \\
Apr 18 & 21:33:54 & 32.91 & 0.82 & 4x700 & 231 & 4.6 & $-$0.9$\pm$6.0 \\
May 02 & 23:10:51 & 46.97 & 1.17 & 4x700 & 265 & 4.0 & $-$14.8$\pm$5.0 \\
May 03 & 23:32:45 & 47.99 & 1.20 & 4x700 & 220 & 5.7 & $-$11.5$\pm$6.6 \\
May 04 & 21:30:12 & 48.90 & 1.22 & 4x700 & 229 & 4.8 & $-$14.8$\pm$6.0 \\
May 11 & 20:34:43 & 55.86 & 1.39 & 4x700 & 190 & 6.5 & $-$12.3$\pm$7.4 \\
May 16 & 20:45:50 & 60.87 & 1.52 & 4x700 & 270 & 3.9 & $-$23.1$\pm$5.0 \\
May 17 & 20:46:53 & 61.87 & 1.54 & 4x700 & 188 & 5.8 & $-$19.6$\pm$7.4 \\
May 20 & 21:13:31 & 64.89 & 1.62 & 4x700 & 210 & 5.9 & $-$21.9$\pm$6.7 \\
May 23 & 20:54:26 & 67.88 & 1.69 & 4x700 & 228 & 5.1 & $-$17.4$\pm$6.1 \\
Jun 02 & 20:58:28 & 77.88 & 1.94 & 4x700 & 214 & 5.2 & $-$9.1$\pm$6.5 \\
Jun 04 & 21:10:44 & 79.89 & 1.99 & 4x700 & 204 & 5.6 & $-$2.7$\pm$6.8 \\
Jun 07 & 20:59:19 & 82.88 & 2.07 & 4x700 & 265 & 3.9 & $-$11.3$\pm$5.1 \\
Jun 08 & 21:04:34 & 83.88 & 2.09 & 4x700 & 278 & 3.8 & $-$6.5$\pm$4.7 \\
\hline
\end{tabular}
\tablefoot{The columns are: (1 and 2) date and universal time of the observations, (3) Heliocentric Julian Date normalised to the first collected observation, (4) rotational cycle of the observations found using Eq.~\ref{eq:ephemeris}, (5) exposure time of a polarimetric sequence, (6) signal-to-noise ratio at 1650 nm per polarimetric sequence, (7) RMS noise level of Stokes $V$ signal in units of unpolarised continuum, (8) longitudinal magnetic field with formal error bar.}
\end{table*}

Characterising the space weather for a specific system and modelling the interaction between the magnetised stellar wind and a close-in planet requires robust knowledge of the stellar magnetic field \citep{Vidotto2014b,Vidotto2014a}. Our assumptions on its topology and strength indeed impact the extent of the planetary magnetosphere \citep{Villareal2018,Carolan2021} and predictions of transits duration \citep{Llama2013}. Stellar magnetic fields are most effectively studied using spectropolarimetry, with which we can analyse the Zeeman effect, i.e. the splitting of spectral lines in distinct components characterised by specific polarisation properties \citep{Zeeman1897}. From time series of polarised spectra we can map the large-scale magnetic field by means of Zeeman-Doppler Imaging (ZDI; \citealt{Semel1989,DonatiBrown1997}) and obtain a global picture of the magnetic environment. Zeeman-Doppler Imaging has been applied extensively in spectropolarimetric studies, and revealed a variety of field geometries for low-mass stars \citep[e.g.,][]{Petit2005,Donati2008,Morin2008,Morin2010,Fares2013,Fares2017}.

GJ~436 is a quiet M2.5 dwarf and hosts a hot-Neptune at 0.0285\,AU, corresponding to an orbital period of 2.644\,d \citep{Butler2004,Gillon2007}. The planet mass is 0.07\,M$_{\rm Jup}$, which places it at the lower-mass boundary of the Neptune desert. The vicinity of the planet to the host star makes it an excellent laboratory to study interactions between the planetary atmosphere and the impinging stellar wind \citep{Vidotto2017,Khodachenko2019,Villareal2021}. Indeed, because of intense irradiation, the planetary atmosphere is subject to hydrodynamic escape, which form a comet-like cloud of hydrogen atoms \citep{Kulow2014,Ehrenreich2015,Lavie2017,DosSantos2019}. To explain the such observations of the system, \citet{Bourrier2015} and \citet{Bourrier2016} showed that an accurate description of the interactions between the stellar wind and the exospheric cloud, together with radiation pressure, is necessary. For instance, variations occurring locally in the cloud structure can be correlated to changes in stellar wind density. The wind properties of the host star GJ\,436 are also important to predict the flux of energetic particles penetrating the atmosphere of GJ\,436\,b \citep{Mesquita2021,RodgersLee2023}.


GJ\,436\,b lies on a polar eccentric orbit \citep{Bourrier2018,Bourrier2022} to which it may have migrated via interactions with an undetected outer companion \citep{Beust2012,Bourrier2018}. The migration would have occurred late in the life of the planet, implying that the latter would have avoided the strong irradiation of the young star and started evaporating only recently and not substantially. This possibly explains why the planet falls in the Neptune desert, but its atmosphere has not been eroded yet \citep{Attia2021}, and represents an interesting case to follow-up. Moreover, depending on the topology of the stellar magnetic field, the planet orbit could sweep regions of both open and closed field lines, as well as oscillate in and out of the Alfv\'en surface, driving intermittent star-planet interactions similar to those modelled for AU\,Mic 
\citep{Kavanagh2021}. The imprints of such interactions would be observable at radio wavelengths \citep[e.g.,][]{Zarka1998,Saur2013,Turnpenney2018, Kavanagh2022}.

In this first paper, we characterise the large-scale magnetic field of GJ\,436 using ZDI on optical spectropolarimetric observations. In a second paper (Vidotto et al., subm.), we will model self-consistently the stellar wind to provide more realistic constraints on the stellar environment at the orbit of GJ\,436\,b. In Sec.~\ref{sec:observations} we describe the spectropolarimetric time series collected with Narval, and we outline the longitudinal field computation and its temporal analysis in Sec.~\ref{sec:Blon} and Sec.~\ref{sec:GP}. The large-scale magnetic field reconstruction by means of Zeeman-Doppler Imaging is presented in Sec.\ref{sec:ZDI}. In Sec.~\ref{sec:conclusion} we summarise and contextualise our results.

\section{Observations}\label{sec:observations}

GJ~436 is an M2.5~dwarf at a distance of 9.76$\pm$0.01 pc \citep{GaiaEDR3} and with a $V$ band magnitude of 10.61 \citep{Zacharias2012}. The stellar radius is 0.417$\pm$0.008\,$R_\odot$ and the mass is 0.441$\pm$0.009\,$M_\odot$ \citep{Rosenthal2021}, placing it above the fully convective boundary at 0.35\,$M_\odot$ \citep{Chabrier1997}. The star is moderately inactive, with a stellar rotation period around 40-44\,d \citep{Bourrier2018,DosSantos2019,Kumar2023} and a chromospheric activity index $\mathrm{logR'}_\mathrm{HK}$ of -5.1 \citep{BoroSaikia2018,Fuhrmeister2023}.

In this work, we used sixteen Narval observations of GJ\,436 collected between March and June 2016 (PI E. Hebrard). The time series is provided in Table~\ref{tab:log}. Narval is the optical spectropolarimeter on the 2~m T\'elescope Bernard Lyot (TBL) at the Pic du Midi Observatory in France, and covering 360-1050\,nm spectral range at a resolving power $R$ of 65 000 \citep{Donati2003}. The data reduction was performed with \texttt{LIBRE-ESPRIT} \citep{Donati1997}, and the reduced spectra were retrieved from PolarBase \citep{Petit2014}. 

From the time series of unpolarised and circularly polarised spectra, we computed high signal-to-noise ratio (S/N) Stokes $I$ and $V$ profiles by means of Least-Square Deconvolution (LSD) \citep{Donati1997,Kochukhov2010}. This numerical technique combines the information of thousands of absorption lines in the observed spectrum, which are selected using a theoretical line list with associated properties such as depth, sensitivity to Zeeman effect (Land\'e factor, $g_\mathrm{eff}$), and excitation potential. 

Considering that GJ\,436 is an M2.5 star with an effective temperature of $3586.1\pm36.4$\,K \citep{Rosenthal2021}, we adopted a line list corresponding to a MARCS model characterised by $\log g=$ 5.0\,[cm s$^{-2}$], $v_{\mathrm{micro}}=$ 1\,km s$^{-1}$, and $T_{\mathrm{eff}}=3500$\,K \citep{Gustafsson2008}. The line list was generated with the Vienna Atomic Line Database\footnote{\url{http://vald.astro.uu.se/}} \citep[VALD,][]{Ryabchikova2015}, and contained 3240 lines in range 350-1080 nm and with depths larger than 40\% the continuum level, similarly to \citet{Morin2008} and \citet{Bellotti2021}. The number of lines takes into account the removal of the following wavelength intervals, that may be affected by residuals of telluric correction or are in the vicinity of H$\alpha$: [627,632], [655.5,657], [686,697], [716,734], [759,770], [813,835], and [895,986] nm.

\begin{figure}[!t]
    \centering
    \includegraphics[width=\columnwidth]{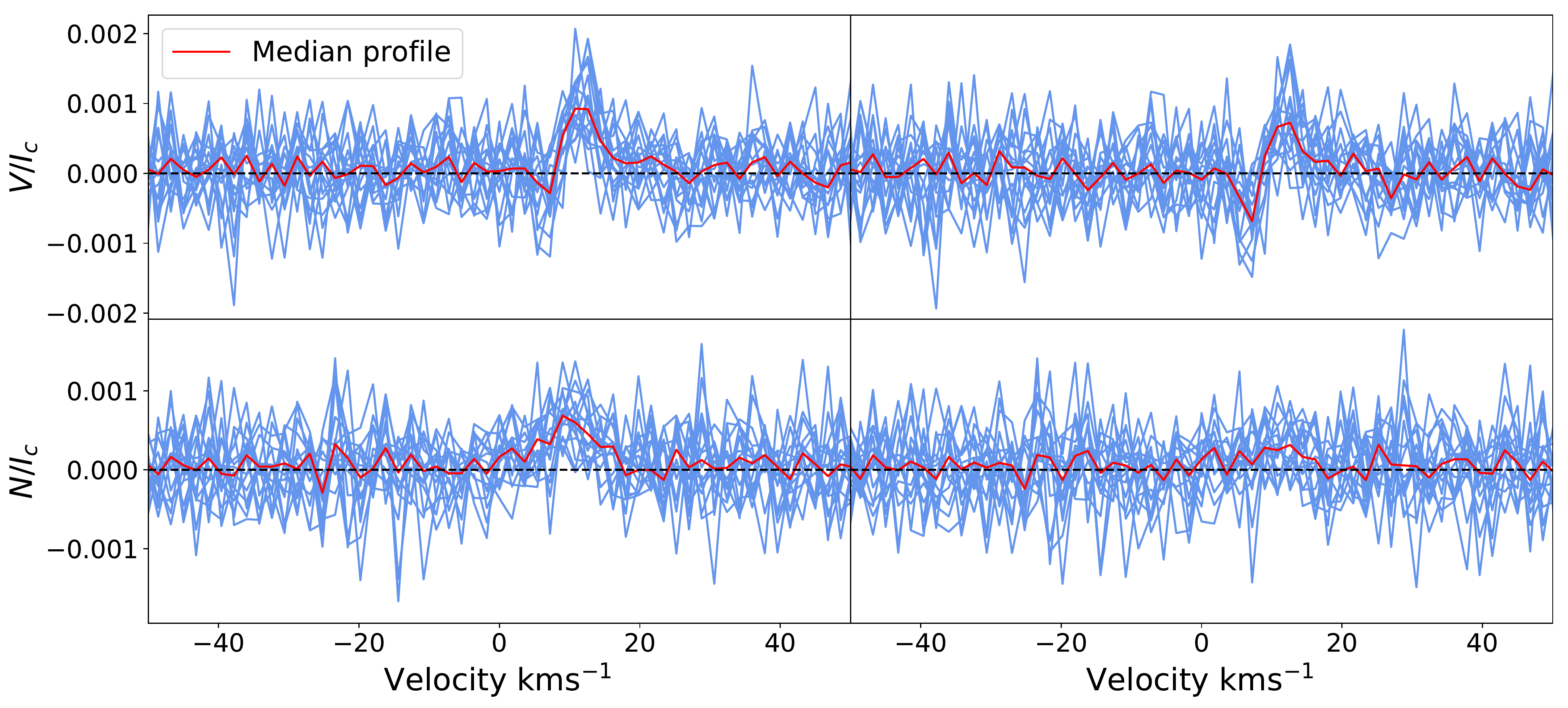}
    \caption{Circular polarisation and Null profiles for the 2016 Narval observations. Left: Stokes~$V$ (top) and Null profile (bottom) profile computed using the full line list between 360-1080 nm. Right: same profiles but obtained using only red ($>$500\,nm) lines. In the latter case, we note that the spurious signal at line centre is removed. The red solid lines in all panels indicate the median profile.}
    \label{fig:Stokes}%
\end{figure}

Along with Stokes $I$ and $V$ profiles, we computed the ``Null profile'', which is a powerful diagnostic tool to determine the noise level of the LSD output and whether a spurious polarisation signal is present in the observations \citep{Donati1997,Bagnulo2009}. As shown in Fig.~\ref{fig:Stokes}, the Null profile contains a positive signal at line centre ($\sim$9.6\,km\,s$^{-1}$), and reflects in a vertical offset of Stokes~$V$ with respect to a constant null value. Following \citet{Folsom2016}, we attributed this signal to an imperfect background subtraction affecting the blue orders of Narval and we removed it by computing LSD profiles using lines in the red part of the spectrum, i.e. larger than 500\,nm. Considering a window of $\pm$10\,km\,s$^{-1}$ from line centre that includes both lobes of the Stokes~$V$ profile, the mean and standard deviation of the Null profile decrease from $3.2\cdot10^{-4}$ to $1.5\cdot10^{-4}$ and from $2.1\cdot10^{-4}$ to $1.3\cdot10^{-4}$, respectively. This procedure does not alter the shape of the Stokes~$V$ profiles, and removes the vertical offset (see Fig.~\ref{fig:Stokes}). The S/N of the final profiles ranges between 1600 and 2600. 

In the following, the observations are phased according to the ephemeris
\begin{align}
    \mathrm{HJD}= 2457464.4967 + \mathrm{P}_\mathrm{rot}\cdot n_\mathrm{cyc}
    \label{eq:ephemeris}
\end{align}
where we used the first collected observation date as reference, P$_\mathrm{rot}$ is the stellar rotation period found using ZDI (see Sec.~\ref{sec:ZDI}), and $n_\mathrm{cyc}$ is the rotation cycle.


\section{Longitudinal magnetic field}\label{sec:Blon}

The longitudinal field (B$_l$) is sensitive to the appearance of magnetic regions on the visible stellar hemisphere, which is modulated by the stellar rotation period (P$_\mathrm{rot}$). As a result, we can generally apply a standard periodogram analysis to B$_l$ time series in order to find P$_\mathrm{rot}$ \citep{Hebrard2016,Petit2021,Carmona2023}.

\begin{figure}[!t]
    \centering
    \includegraphics[width=\columnwidth]{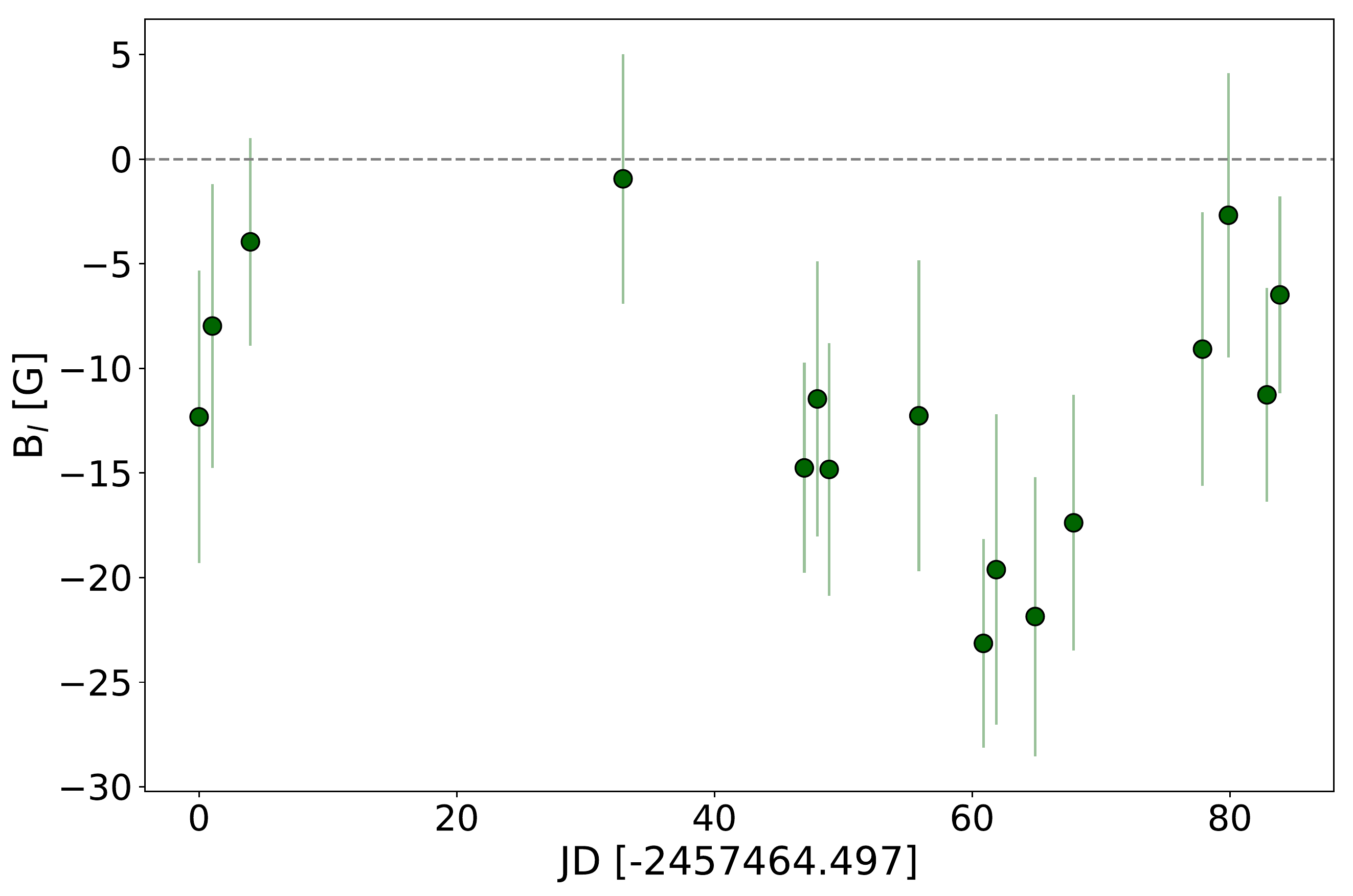}
    \includegraphics[width=\columnwidth]{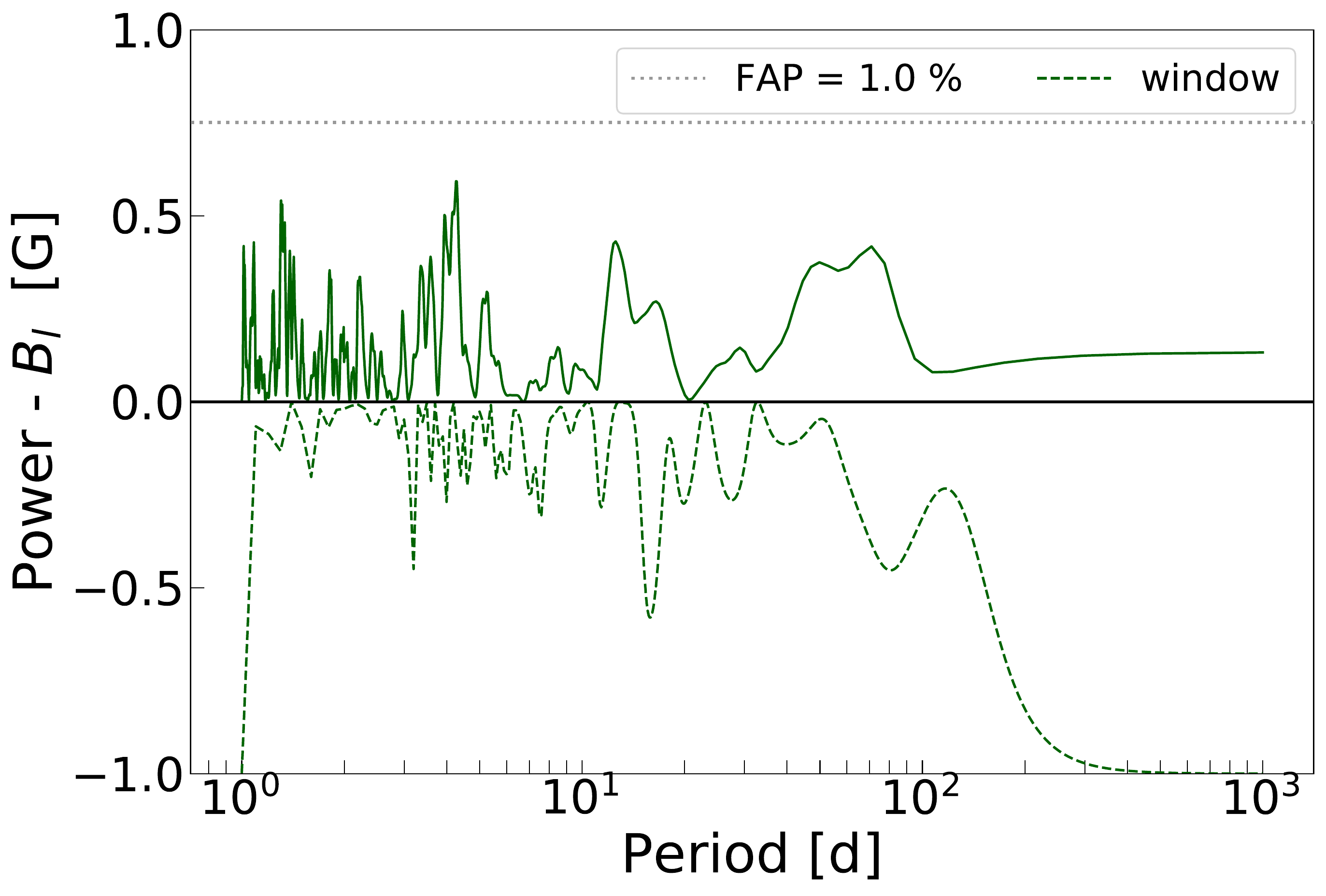}
    \caption{Analysis of longitudinal field measurements. Top: Time series of B$_l$ measurements. All values have a negative sign, and range between $-$1 and $-$23\,G. Bottom: generalised Lomb-Scargle Periodogram of the longitudinal field time series. The analysis does not yield any significant (FAP$<$1\%) periodicity. The window function of the entire time series is included and mirrored with respect to the x axis \citep{VanderPlas2018}, to highlight aliases due to the observation cadence.}
    \label{fig:Bl_LSP}%
\end{figure}

Previous studies extracted a stellar rotation period of 39.9$\pm$0.8\,d from chromospheric activity indexes time series \citep{SuarezMascareno2015,DosSantos2019} and 44.09$\pm$0.08 d from photometric data sets \citep{Bourrier2018}. Recently, \citet{Kumar2023} analysed GJ\,436's spectra obtained with HARPS and Narval and, by computing time series of activity indexes such as Ca\textsc{II} and H$\alpha$, found a significant (the false-alarm probability, i.e. FAP, was less than 0.1\%) periodogram peak at $39.47^{+0.11}_{-0.15}$ and $40.46^{+0.44}_{-0.52}$\,d, respectively. The Narval data set used by \citet{Kumar2023} was the same one employed in this work.

We followed \citet{Donati1997} to compute the disk-averaged, line-of-sight-projected stellar magnetic field as the first-order moment of a Stokes~$V$ profile
\begin{equation}
\mathrm{B}_l\;[G] = \frac{-2.14\cdot10^{11}}{\lambda_0 \mathrm{g}_{\mathrm{eff}}c}\frac{\int vV(v)dv}{\int(I_c-I)dv} \,
\label{eq:Bl}
\end{equation}
where $\lambda_0$ (in nm) and $\mathrm{g}_\mathrm{eff}$ are the normalisation wavelength and Land\'e factor of the LSD profiles, $I_c$ is the continuum level, $v$ is the radial velocity in the star's rest frame and $c$ the speed of light in vacuum (both in km\,s$^{-1}$).

We used a normalisation wavelength and Land\'e factor of 700\,nm and 1.1976, respectively, and performed the integration within $\pm$10\,km\ s$^{-1}$ from line centre at around 9.6\,km\ s$^{-1}$. The B$_l$ time series is illustrated in Fig.~\ref{fig:Bl_LSP}, with all values featuring a negative sign.
The mean value is $-$12\,G and both the dispersion and mean error bar are 6\,G.

Fig.~\ref{fig:Bl_LSP} shows the application of a Generalised Lomb-Scargle periodogram \citep{Zechmeister2009} to the entire 2016 time series. We do not report any dominant periodicity, the FAP being systematically higher than 1\%. The highest peaks are around 4, 15, and 70\,d, but they are probably generated by the sparse sampling of our observations, as illustrated by the window function. 

\section{Gaussian Process Regression}\label{sec:GP}

We performed a quasi-periodic Gaussian Process (GP; \citealt{Haywood2014}) fit to the longitudinal field curve, since this model is more flexible than the standard sine function used in the Lomb-Scargle analysis. In fact, the GP model accounts for the evolution of the magnetic field and its variability \citep{Aigrain2022}. Formally, we used the quasi-periodic covariance function
\begin{equation}\label{eq:qp_kernel}
    k(t,t') = \theta_1^2\exp\left[-\frac{(t-t')^2}{\theta_2^2}-\frac{\sin^2\left(\frac{\pi(t-t')}{\theta_3}\right)}{\theta_4^2} \right] + S^2\delta_{t,t'},
\end{equation}
where $\delta_{t,t'}$ is a Kronecker delta, and $\theta_i$ are the hyperparameters of the model: $\theta_1$ is the amplitude of the curve in G, $\theta_2$ is the evolution timescale in d (it expresses how rapidly the model evolves), $\theta_3$ is the recurrence timescale (i.e., P$_\mathrm{rot}$), and $\theta_4$ is the smoothness factor (controlling the harmonic structure of the curve). We added an additional hyperparameter to account for the excess of uncorrelated noise ($S$). In practice we used the \textsc{cpnest} package \citep{DelPozzo2022} which performs Bayesian inference via nested sampling algorithm \citep{Skilling2004}.

\begin{table}[!t]
\caption{Results of the GP fit carried out on the B$_l$ curve of GJ\,436.}
\label{tab:gp}
\centering
\begin{tabular}{l l l} 
\hline
Hyperparameter & Prior & Best fit value\\
\hline
Amplitude [G] ($\theta_1$) & $\mathcal{U}(0,100)$ & $16.4^{+13.0}_{-6.2}$\\
Decay time [d] ($\theta_2$) & $\mathcal{U}(1,1000)$ & $440^{+370}_{-310}$\\
P$_\mathrm{rot}$ [d] ($\theta_3$) & $\mathcal{U}(1,60)$ & $46.6^{+4.8}_{-6.8}$\\
Smoothness ($\theta_4$) & $\mathcal{U}(0.1,1.2)$ & $0.9^{+0.2}_{-0.3}$\\
Uncorrelated noise [G] ($S$) & $\mathcal{U}(0,100)$ & $2.7^{+1.8}_{-1.6}$\\
\hline
\end{tabular}
\tablefoot{The columns are: (1) hyperparameter, (2) prior uniform distribution of the form $\mathcal{U}(\mathrm{min},\mathrm{max})$, and (3) median of the posterior distribution with 16th and 84th percentiles error bars.}
\end{table}

\begin{figure}[!t]
    \centering
    \includegraphics[width=\columnwidth]{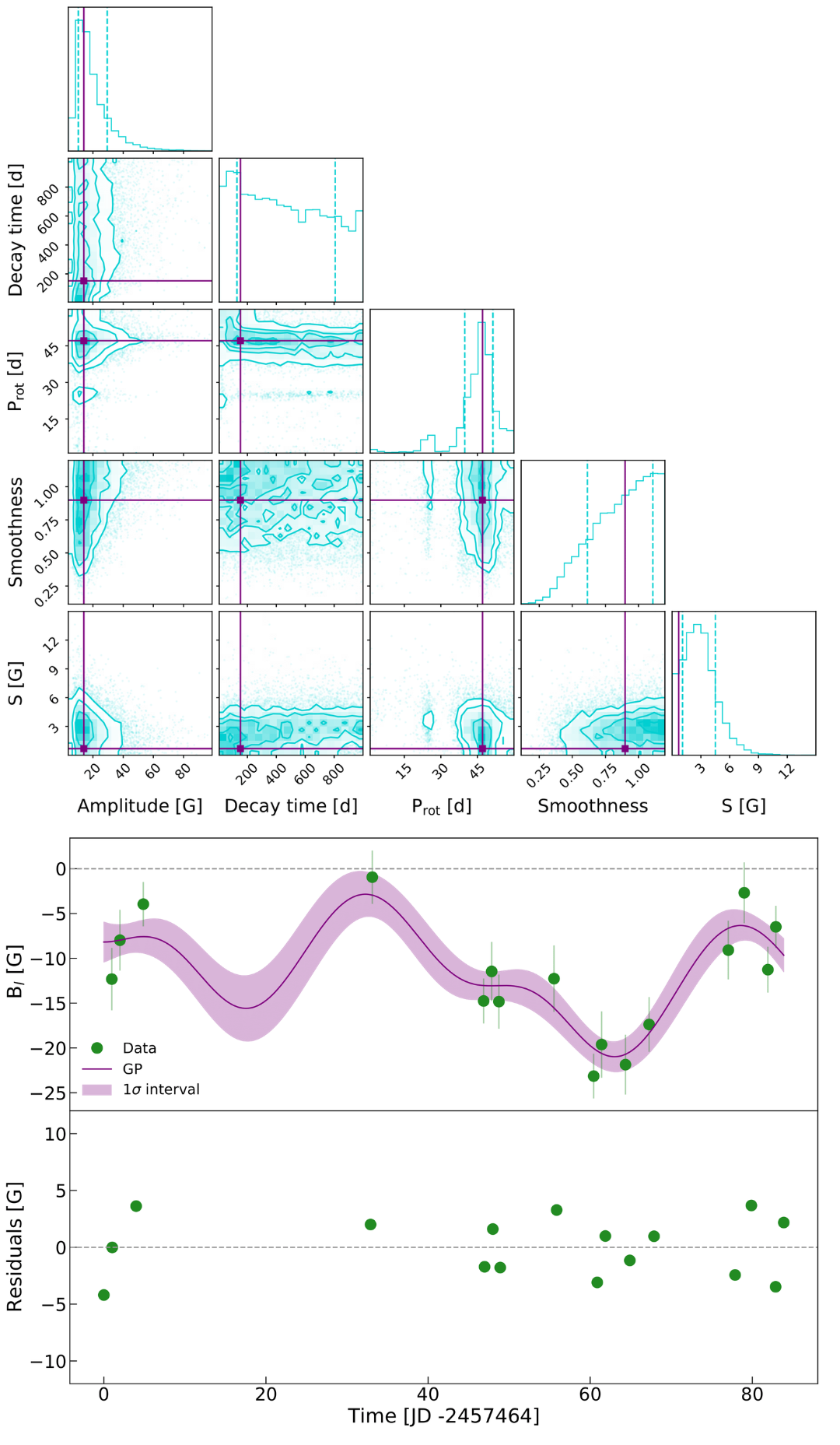}
    \caption{Gaussian Process Regression applied to longitudinal field. Top: corner plot display the 2D posterior distributions of the hyperparameters of the GP model (see Eq.~\ref{eq:qp_kernel}) as well as the 1D marginalised distributions along the diagonal. Vertical solid lines indicate the median of the distribution, while dashed lines indicate the 16th and 84th percentiles. Bottom: GP model overplotted to the time series of B$_l$ values and residuals of the model. The shaded area indicates the 1$\sigma$ uncertainty region.}
    \label{fig:gp}%
\end{figure}

The results are reported in Table~\ref{tab:gp} and showed in Fig.~\ref{fig:gp}. We applied uniform priors to all five hyperparameters, and allowed the search within realistic boundaries. The model fits the data to a $\chi^2_r$ of 0.6, likely indicating that our formal error bars are overestimated. Following \citet{Donati2023}, we rescaled the error bars by a factor of 2 to fit the data at $\chi^2_r=1.0$, while keeping the excess of uncorrelated noise consistent with zero (see Fig.~\ref{fig:gp}). The GP model is characterised by smooth oscillations (i.e., $\theta_4$=1.1), with an amplitude of 12\,G and a stellar rotation period of 46.6\,d, which is in agreement with the value estimated in the literature within error bars \citep{SuarezMascareno2015,Bourrier2018,Kumar2023}. The dispersion of the residuals is 2.5\,G, i.e. slightly lower than the rescaled error bars.

For P$_\mathrm{rot}$, using a uniform prior between 1 and 100\,d results in a marginalised posterior distribution with three peaks, around 20\,d, 40\,d and 80\,d, with the latter being the highest peak. From the Lomb-Scargle analysis presented in Fig.~\ref{fig:Bl_LSP}, we observe that the observing window function features a broad peak at 80\,d, hence we can exclude it from being the genuine rotation period of the star. If we lower the uniform prior boundary to 60\,d, the marginalised posterior distribution exhibits a maximum at 46.6\,d.

We also notice that the evolution time scale $\theta_2$ is not constrained by the GP. This is not surprising given the short (i.e. 80\,d) time span of our observations. We therefore fixed $\theta_2$ to either 200 or 300\,d, following the results of starspots lifetime analysis carried out by \citet{Giles2017}, and performed a 4-hyperparameters GP fit, but the results were only marginally different than those obtained with a 5-hyperparameters GP. We also attempted an analogous test fixing a decay time of 470\,d, i.e. the active regions timescale reported by \citet{Kumar2023}, but the results did not differ. Finally, a similar conclusion is obtained when fixing both the decay time scale and the smoothness to the values constrained by \citet{Martioli2022} for TOI-1759, which is an M~dwarf of similar spectral type as GJ~436. We used 400-600\,d and 0.7-0.9 for $\theta_2$ and $\theta_4$, respectively. 

Finally, although the GP retrieves the stellar rotation period around the expected value, we note that the error bars of such time scale are large. An alternative option to extract the stellar rotation period is via ZDI optimisation, as outlined in the next section.

\section{Zeeman-Doppler Imaging}\label{sec:ZDI}

\begin{figure}[!t]
    \centering
    \includegraphics[width=\columnwidth]{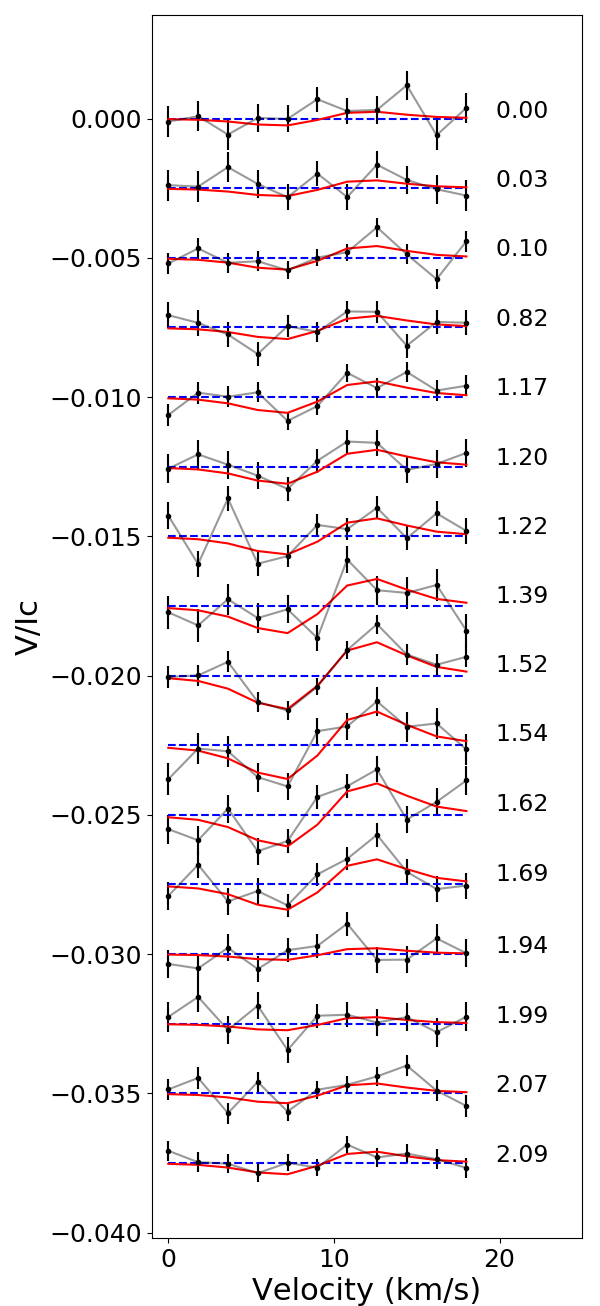}
    \caption{Narval time series of circularly polarised Stokes profiles. Observations are shown as black dots and ZDI models as red lines, and are offset vertically for better visualisation. The number on the right indicates the rotational cycle (see Eq.~\ref{eq:ephemeris}). All signatures are antisymmetric, indicating that we are seeing the negative polarity of a dipole, and the moderate variation in amplitude is symbolic of a small tilt of the magnetic axis.}
    \label{fig:stokesV}%
\end{figure}

We reconstructed the large-scale magnetic field at the surface of GJ\,436 by means of ZDI. The field is formally described as the sum of a poloidal and toroidal component, both expressed via spherical harmonic decomposition \citep{Donati2006, Lehmann2022}. With ZDI, we synthesise and adjust Stokes~$V$ profiles in an iterative fashion, until a maximum-entropy solution at a fixed reduced $\chi^2$ is achieved \citep{Skilling1984,DonatiBrown1997,Folsom2018}. The iterative process aims to fit the spherical harmonics coefficients $\alpha_{\ell,m}$, $\beta_{\ell,m}$, and $\gamma_{\ell,m}$ (with $\ell$ and $m$ the degree and order of the mode, respectively).

We optimised the input stellar rotation period following the method described in \citet{Petit2002} and \citet{Morin2008}. Basically, we sought the value minimising the $\chi^2_r$ distribution at a fixed entropy (information content) over a grid of possible values between 2 and 100\,d. We found P$_\mathrm{rot}=40.13\pm1.29$\,d, which is compatible with the GP model estimate in Sec.~\ref{sec:GP}, as well as with literature estimates \citep{Bourrier2018,Kumar2023}. For the other input parameters, we adopted an inclination of 40$^{\circ}$ and an equatorial projected velocity ($v_e\sin(i)$) of 0.33\,km\,s$^{-1}$ \citep{Bourrier2022}. We further assumed solid body rotation, a linear limb darkening law with a $V$-band coefficient of 0.6964 \citep{Claret2011}, and the maximum degree of harmonic expansion $\ell_\mathrm{max}=5$, to match the spatial resolution determined by the $v_e\sin(i)$ of the star. The Narval Stokes~V time series is shown in Fig.~\ref{fig:stokesV}.




\begin{figure}[!t]
    \centering
    \includegraphics[width=\columnwidth]{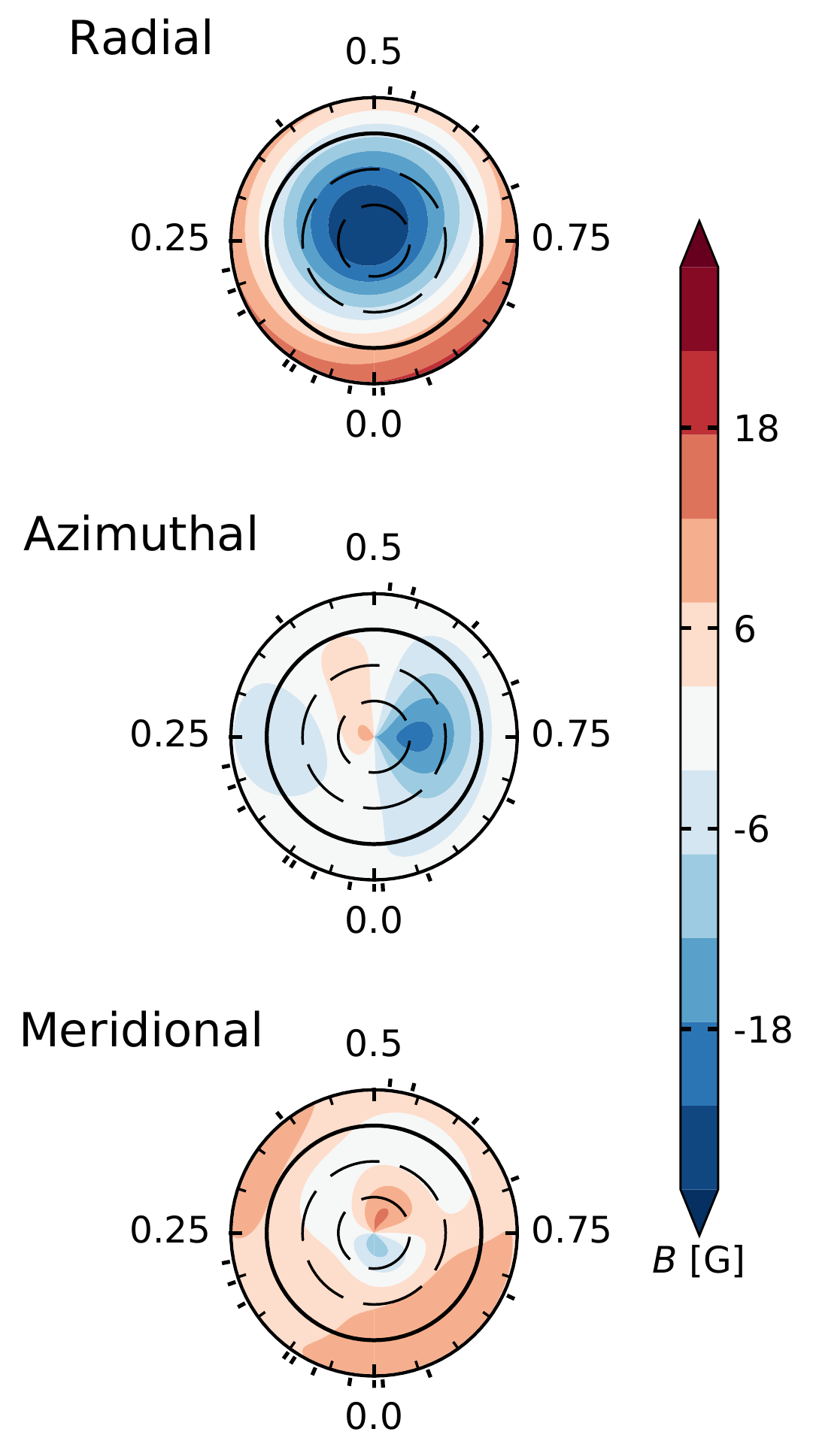}
    \caption{Zeeman-Doppler Imaging reconstruction in flattened polar view of the large-scale field of GJ\,436. From the top, the radial, azimuthal and meridional components of the magnetic field vector are displayed. The radial ticks are located at the rotational phases when the observations were collected, while the concentric circles represent different stellar latitudes: -30\,$^{\circ}$, +30\,$^{\circ}$ and +60\,$^{\circ}$ (dashed lines) and equator (solid line). The geometry is predominantly poloidal, dipolar and axisymmetric. The colour bar encapsulates the magnetic field strength, up to a maximum of 31\,G.}
    \label{fig:zdi}%
\end{figure}

The model Stokes~$V$ profiles are fit down to $\chi^2_r=1.16$, from an initial value of 2.24. The target $\chi^2_r$ represents the best value that avoids underfitting and overfitting of the Stokes~$V$ shape, resulting in a, e.g., weaker field or spurious magnetic features, respectively. The magnetic map is illustrated in Fig.~\ref{fig:zdi} and its properties are listed in Table~\ref{tab:zdi_output}. The mean magnetic field strength is B$_\mathrm{mean}=$ 16\,G, with the poloidal component accounting for 96\% of the magnetic energy. The dipolar and quadrupolar modes store 90\% and 8\% energy, and the field is mostly axisymmetric (79\%), with an obliquity of its axis of 15.5$^{\circ}$.

Zeeman Doppler Imaging does not provide error bars on the reconstructed maps, and thus on field characteristics. We estimated variation bars on the field characteristics following the method of \citet{Mengel2016} and \citet{Fares2017}. We reconstructed magnetic maps for the input parameters (inclination, $v_\mathrm{eq}\sin(i)$, and P$_\mathrm{rot}$) by varying each of them within their error bars. The variation bars reported in Table~\ref{tab:zdi_output} correspond to the maximum difference of field characteristics between the map with the optimised set of input parameters, and the ones considering the error bars on the input parameters. 
We also reconstructed the magnetic field topology using P$_\mathrm{rot}$ = 44.09\,d as input \citep{Bourrier2018}. The target $\chi^2_r$ was adjusted to a larger value of 1.18, but the final map was consistent with the one presented in Fig.~\ref{fig:zdi} within variation bars.


\begin{table}[!t]
\caption{Properties of the magnetic map.} 
\label{tab:zdi_output}     
\centering                       
\begin{tabular}{l r}      
\hline     
B$_\mathrm{mean}$ [G]   & 15.9$^{+0.8}_{-1.5}$\\
B$_\mathrm{max}$ [G]    & 30.9$^{+4.5}_{-0.5}$\\
B$_\mathrm{pol}$ [\%]   & 96.4$^{+0.6}_{-4.0}$\\
B$_\mathrm{tor}$ [\%]   & 3.6$^{+3.6}_{-0.6}$\\
B$_\mathrm{dip}$ [\%]   & 90.4$^{+0.9}_{-11.6}$\\
B$_\mathrm{quad}$ [\%]  & 7.8$^{+8.9}_{-0.8}$\\
B$_\mathrm{oct}$ [\%]   & 1.7$^{+2.5}_{-0.2}$\\
B$_\mathrm{axisym}$ [\%]& 78.7$^{+1.8}_{-15.7}$\\
Obliquity [$^{\circ}$] & 15.5$^{+2.0}_{-2.0}$\\
\hline                                 
\end{tabular}
\tablefoot{The following quantities are listed: mean magnetic strength, maximum magnetic strength, poloidal and toroidal magnetic energy as a fraction of the total one, dipolar, quadrupolar and octupolar magnetic energy as a fraction of the poloidal one, axisymmetric magnetic energy as a fraction of the total one, and tilt of the magnetic axis relative to the rotation axis. The variation bars are computed by reconstructing ZDI maps including the uncertainties on the input stellar parameters (see text).}
\end{table}

\begin{figure}[!ht]
    \centering
    \includegraphics[width=\columnwidth]{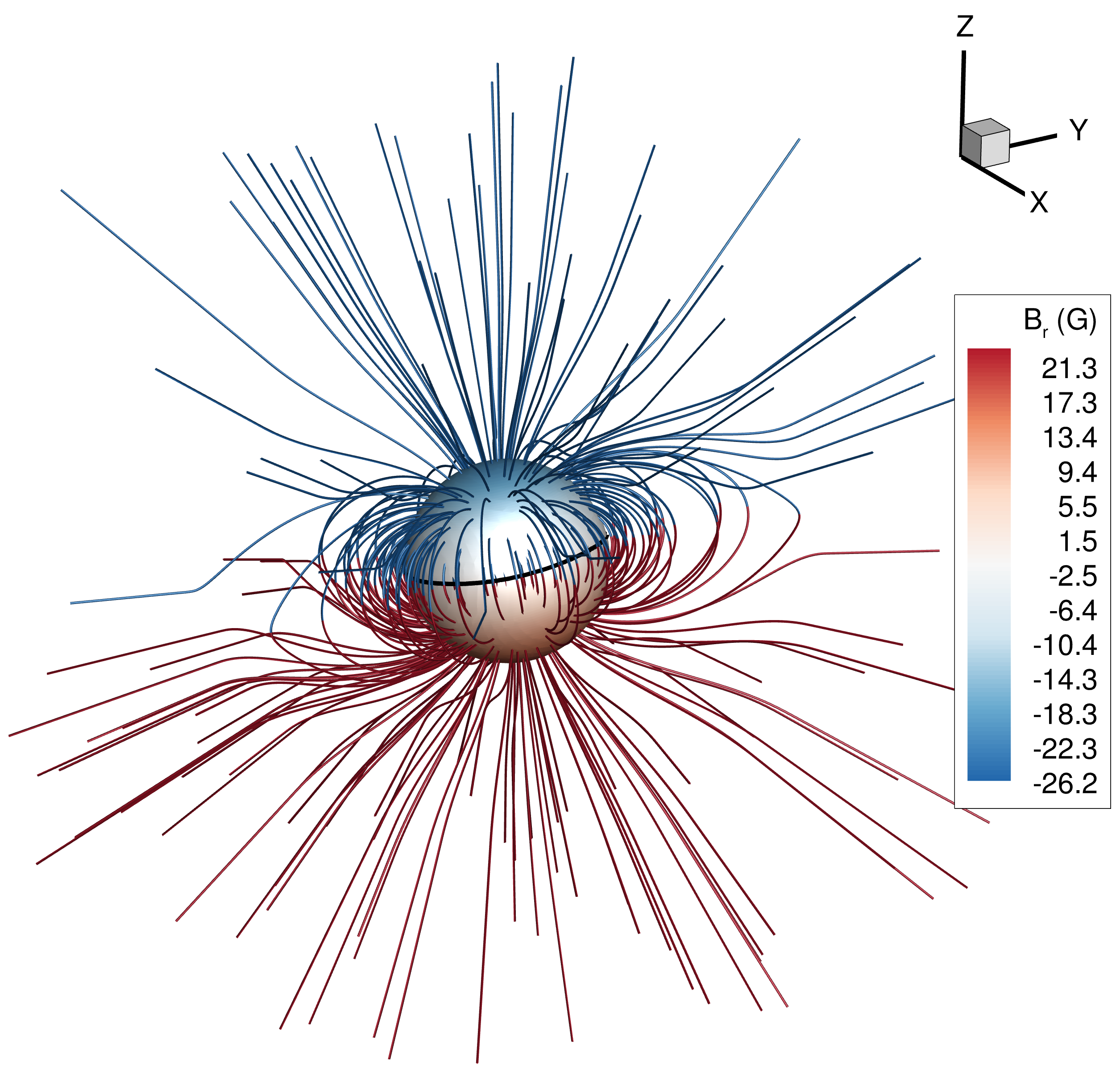}
    \caption{Three dimensional view of the extrapolated large-scale magnetic field of GJ~436. The colours at the surface of the star represent the radial magnetic field strength, while the blue/red colours along the magnetic field lines represent negative/positive polarities of the radial field. The rotation axis of the star is along the $Z$-axis and the source surface is set to 4 stellar radii, beyond which the field lines are fully open.}
    \label{fig:potential}%
\end{figure}

For illustration, Fig.~\ref{fig:potential} shows an extrapolation of the surface field of the star. We used a potential field source surface method \citep[e.g.][]{Jardine2002}, adopting a source surface at a distance of 4 stellar radii -- beyond this distance, the field lines are fully open and purely radial. Using this extrapolation method, we found that at the orbital distance of GJ\,436\,b (0.028\,au; \citealt{Butler2004}), the radial magnetic field ranges from $-0.050 ^{+0.010} _{-0.002}$\,G to $0.048^{+0.002}_{-0.010}$\,G, with the negative value representing an inward radial field and the positive value an outward radial field. 
In a follow up study, we will perform stellar wind modelling and provide more detailed predictions of the characteristics of the wind environment (including its embedded magnetic field) at the orbit of GJ\,436\,b.






\begin{figure*}[!ht]
    \centering
    \includegraphics[width=\textwidth]{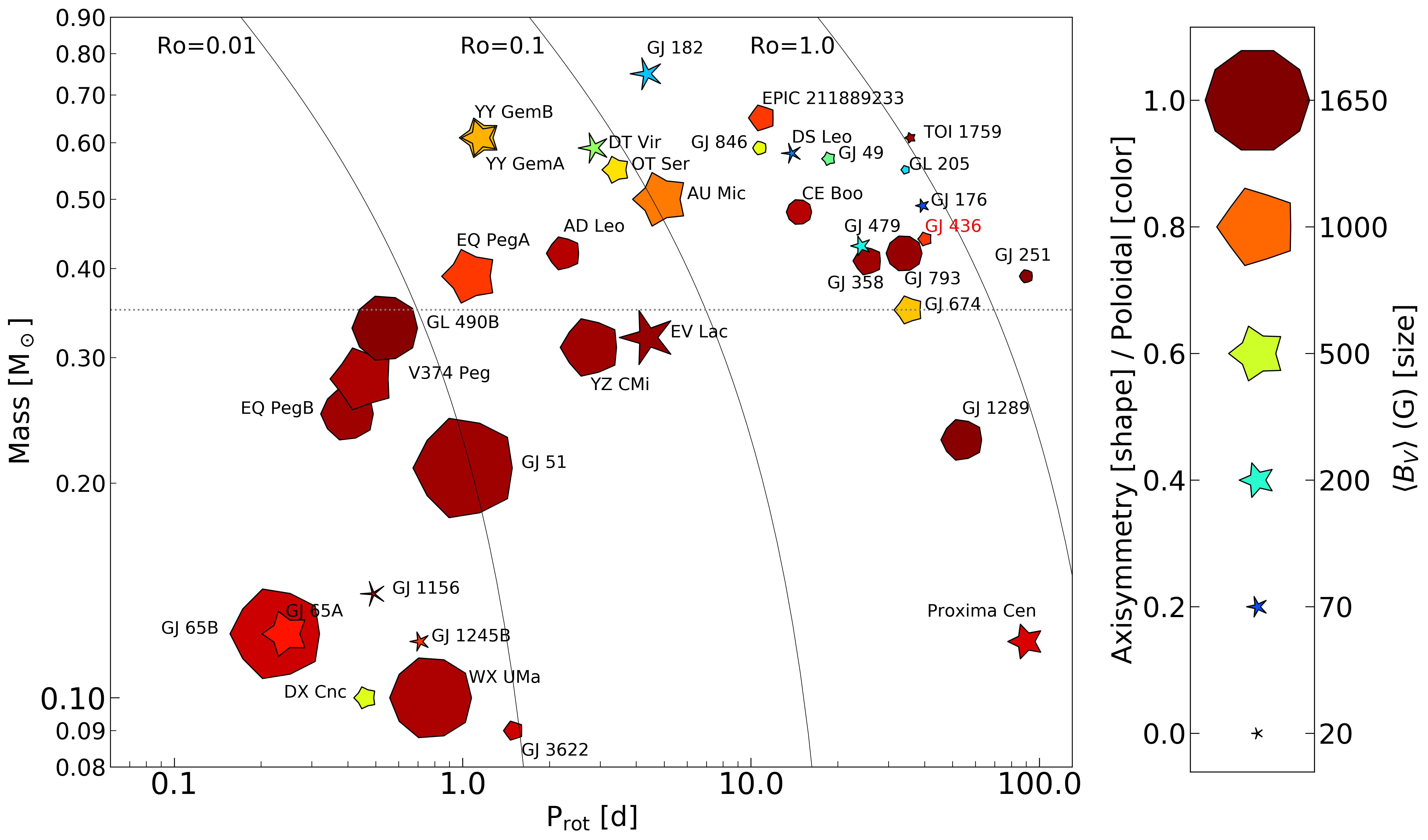}
    \caption{Properties of the magnetic topologies for cool, main-sequence stars obtained via Zeeman-Doppler Imaging. The label of GJ\,436 is highlighted in red. The $y$- and $x$-axes represent the mass and rotation period of the star, and iso-Rossby number curves are overplotted using the empirical relations of \citet{Wright2018}. 
    The symbol size, colour and shape encodes the ZDI average field strength, poloidal/toroidal energy fraction and axisymmetry. Data entering the plot are taken from \citealt{Donati2008,Morin2008,Phan-Bao2009,Morin2010,Hebrard2016,Kochukhov2017,Moutou2017,Kochukhov2019,Klein2021,Martioli2022,CortesZuleta2023}.}
    \label{fig:confusogram}%
\end{figure*}

\section{Discussion and conclusions}\label{sec:conclusion}

In this paper, we presented the analysis of the large-scale magnetic field of the exoplanet host star GJ~436. This will serve as input for the stellar wind and star-planet interaction analysis which will be presented in a future paper (Vidotto et al., subm.). The main goal is to understand stellar environments around M~dwarfs, which is relevant for both exoplanet searches and habitability assessment frameworks \citep{Vidotto2013, OMalley-James2019,Lingam2019}. Ultimately, this will provide insightful feedback on the influence of stellar magnetic fields on planetary atmospheres and habitability, which is of crucial importance for {\it JWST} and {\it Ariel}, since GJ\,436 is in the reference sample of both missions \citep{Edwards2022}.


We used spectropolarimetric data collected with Narval in 2016, and we computed the longitudinal magnetic field from the time series of circularly polarised spectra. To the same time series, we applied tomographic inversion (i.e., Zeeman-Doppler Imaging) to reconstruct a map of the large-scale magnetic field topology. Our conclusions are summarised as follows:

\begin{enumerate}
    \item The longitudinal field (B$_l$) spans between $-$0.9 and $-$23.1\,G, with a median error bar of 6\,G. Such field strength is comparable with that of other M~dwarfs with similar spectral types and rotation periods.
    \item A periodicity analysis by means of generalised Lomb-Scargle periodogram applied to the B$_l$ time series did not highlight any specific periodicity, similarly to the activity indexes analysis of \citet{Kumar2023}. More specifically, we did not retrieve the expected rotation period of about 40\,d, but observe different insignificant (FAP>1\%) peaks mostly associated with the observational window. We found $TESS$ \citep{Ricker2014} observations of GJ~436 collected in 2020 and 2022, but in both cases the observing window is shorter than the expected rotation period of the star, hence they cannot be used to constrain such parameter.
    \item The GP regression analysis applied to the B$_l$ time series produces a smooth model characterised by a rotation period of $46.6^{+4.8}_{-6.8}$\,d. From the optimisation of stellar input parameters with Zeeman-Doppler Imaging, we were able to infer P$\mathrm{rot}=40.13\pm1.29$\,d. Both values are in agreement with literature estimates within uncertainties.
    \item The application of Zeeman-Doppler Imaging to the Stokes~$V$ time series revealed a simple field configuration, characterised by a poloidal, mainly dipolar and axisymmetric topology, with a mean magnetic field strength of 16\,G. This simple geometry is in accordance with other stars of similar spectral type, mass and rotation period, i.e. GJ\,205 \citep{Hebrard2016, CortesZuleta2023} and TOI-1759 \citep{Martioli2022}, as can be seen in Fig.~\ref{fig:confusogram}.
\end{enumerate}

\begin{figure}[!t]
    \centering
    \includegraphics[width=\columnwidth]{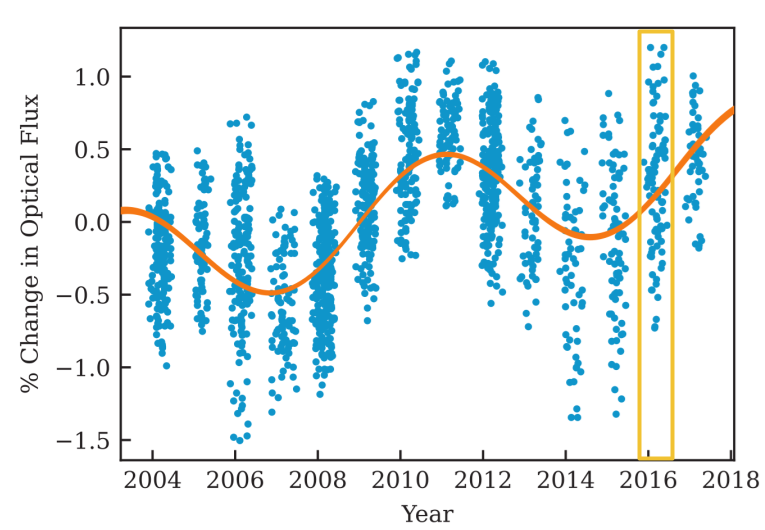}
    \caption{Photometric cycle reported in \citet{Lothringer2018} and \citet{Loyd2023}. Blued data points represent the photometric observations while the orange line represents the sinusoidal fit at a period of 7.75\,yr combined with a linear trend. The yellow box on the right of the plot indicates the time window of our Narval observations in 2016. The figure was adapted form \citet{Loyd2023}.}
    \label{fig:photocycle}%
\end{figure}


GJ\,436 is known to have an activity cycle: \citet{Lothringer2018} analysed 14 years of photometric data (in Str\"omgren $b$ and $y$ filters) between 2004 and 2018, and reported a 7.4\,yr cycle, which was then re-analysed by \citet{Loyd2023} who consistently found a 7.75\,yr cycle. Moreover, a similar time scale between 5 and 7\,yr was obtained by \citet{Kumar2023} from time series of chromospheric activity indexes (H$\alpha$, Na\textsc{I}, and Ca\textsc{II} H\&K), spanning 14\,yr. This in agreement with what is expected for M~dwarfs from radial velocity exoplanet searches \citep{GomesDaSilva2012} and photometric surveys \citep{SuarezMascareno2016,SuarezMascareno2018} for M~dwarfs of similar spectral type.

In this light, it is interesting to place the magnetic field map we reconstructed along the track of the activity cycle. Our observations were collected in 2016 (see Table~\ref{tab:log}), meaning that our ZDI map portrays the magnetic field during an ascending phase of the cycle (i.e. towards photometric maximum), as shown in the yellow box in Fig.~\ref{fig:photocycle}.
This advocates for additional spectropolarimetric monitoring of GJ\,436, in order to ideally reconstruct a ZDI map during cycle minimum and maximum, and catch whether the magnetic field undergoes polarity reversals like for the Sun \citep{Sanderson2003,Lehmann2021}, and other stars and other cool stars (e.g. $\tau$\,Boo \citealt{Fares2009,Mengel2016,Jeffers2018} and 61\,Cyg \citealt{BoroSaikia2016}). If we assume P$_\mathrm{cyc}=7.75$\,yr, we predict the next photometric minimum to be around 2030, whereas the next maximum to be around mid 2026. Monitoring the secular evolution of the large-scale field of GJ~436 would be an essential ingredient to interpret the observed signatures of star-planet interactions. Indeed, magnetic cycles modulate the radiation output of stars \citep{Yeo2014,Hazra2020}, therefore providing a temporal modulation of planetary atmospheric erosion.

\begin{acknowledgements}
We thank the anonymous referee for the fruitful review of this work. We acknowledge funding from the French National Research Agency (ANR) under contract number ANR-18-CE31-0019 (SPlaSH). RF acknowledges support from the United Arab Emirates University (UAEU) startup grant number G00003269. 
This work has been carried out within the framework of the NCCR PlanetS supported by the Swiss National Science Foundation under grants 51NF40$\_$182901 and 51NF40$\_$205606. This project has received funding from the European Research Council (ERC) under the European Union's Horizon 2020 research and innovation programme (projects {\sc Spice Dune} and ASTROFLOW, grant agreements No 947634 and 817540). This work has made use of the VALD database, operated at Uppsala University, the Institute of Astronomy RAS in Moscow, and the University of Vienna; Astropy, 12 a community-developed core Python package for Astronomy \citep{Astropy2013,Astropy2018}; NumPy \citep{VanderWalt2011}; Matplotlib: Visualization with Python \citep{Hunter2007}; SciPy \citep{Virtanen2020} and PyAstronomy \citep{Czesla2019}.

\end{acknowledgements}

%
%

\bibliographystyle{aa}
\bibliography{biblio}

\end{document}